\documentclass[11pt]{iopart}

\newcommand{\bea}{\begin{eqnarray}}
\newcommand{\eea}{\end{eqnarray}}

\begin{document}

\title{Axion as a cold dark matter candidate: Analysis to third order perturbation for classical axion}
\author{Hyerim Noh}
\address{Korea Astronomy and Space Science Institute,
         Daejeon 305-348, Republic of Korea}
\ead{hr@kasi.re.kr}
\author{Jai-chan Hwang}
\address{Department of Astronomy and Atmospheric Sciences,
         Kyungpook National University, Daegu 702-701, Republic of Korea}
\ead{jchan@knu.ac.kr}
\author{Chan-Gyung Park}
\address{Division of Science Education and Institute of Fusion Science,
         Chonbuk National University, Jeonju 561-756, Republic of Korea}
\ead{park.chan.gyung@gmail.com}


\begin{abstract}

We investigate aspects of axion as a coherently oscillating massive classical scalar field by analyzing third order perturbations in Einstein's gravity in the axion-comoving gauge. The axion fluid has its characteristic pressure term leading to an axion Jeans scale which is cosmologically negligible for a canonical axion mass. Our classically derived axion pressure term in Einstein's gravity is identical to the one derived in the non-relativistic quantum mechanical context in the literature. We show that except for the axion pressure term, the axion fluid equations are exactly the same as the general relativistic continuity and Euler equations of a zero-pressure fluid up to third order perturbation. The general relativistic density and velocity perturbations of the CDM in the CDM-comoving gauge are exactly the same as the Newtonian perturbations to the second order (in all scales), and the pure general relativistic corrections appearing from the third order are numerically negligible (in all scales as well) in the current paradigm of concordance cosmology. Therefore, here we prove that, in the super-Jeans scale, the classical axion can be handled as the Newtonian CDM fluid up to third order perturbation. We also show that the axion fluid supports the vector-type (rotational) perturbation from the third order. Our analysis includes the cosmological constant.

\end{abstract}

\tableofcontents

%
%
\section{Introduction}

The currently popular concordance cosmology based on perturbed Friedmann world model demands the presence of substantial amount of non-baryonic dark matter. Despite some problems concerning small-scale (less than some kpc) clustering properties, the cold dark matter (CDM) still plays a popular role as the dark matter. Axion, notwithstanding its still hypothetical nature, is known to be one of the prime candidates for the CDM \cite{Axion-CDM}.

Axion as a coherently oscillating scalar field is known to behave as a pressureless fluid in the background world model \cite{Axion-CDM}, and to the linear order perturbation in the Newtonian context \cite{Khlopov-etal-1985}.
The CDM nature of linear order perturbation of the axion in Einstein's gravity was studied in \cite{Nambu-Sasaki-1990,Ratra-1991,Axion-1997,Sikivie-Yang-2009,Axion-2009}.
The CDM nature of axion to the linear order was shown in various gauge conditions: these are the zero-shear gauge \cite{Nambu-Sasaki-1990,Sikivie-Yang-2009}, the synchronous gauge \cite{Ratra-1991}, the uniform-curvature gauge \cite{Axion-1997}, and the axion-comoving gauge \cite{Axion-2009}. Only in the axion-comoving gauge the axion is shown to behave as the Newtonian CDM in {\it all} cosmological scales \cite{Axion-2009}. The axion, however, has its characteristic pressure term with Solar-System size Jeans scale for the canonical axion mass \cite{Khlopov-etal-1985,Nambu-Sasaki-1990,Sikivie-Yang-2009,Axion-2009}, see Eq.\ (\ref{Jeans-scale}). We may increase the Jeans scale to cosmologically relevant one by reducing the axion mass; roles of this extremely low mass axion as a potential variant of warm dark matter with small scale cut off in the density power spectrum deserve further study \cite{Axion-low-mass-2012}.

Previously we have shown that the axion as a classical massive scalar field behaves as the CDM to the second-order perturbations in all cosmological scales \cite{Axion-second-order}. Except for the axion pressure term relevant in the small-scale limit, the relativistic continuity and Euler equations of the axion fluid are shown to be exactly the same as Newtonian equations to the second order in all cosmological scales including the super-horizon scale; in the case of a zero-pressure and irrotational fluid, see \cite{Noh-Hwang-2004}.

Here, we extend our study to the third-order perturbation in the axion-comoving gauge. The leading nonlinear contribution to the density and velocity power spectra needs the third order perturbation. We will consider the axion as a classical massive scalar field in Einstein's gravity. We will use the fully nonlinear perturbation formulation in \cite{fNL-2013,Noh-2014}. We will show that the general relativistic CDM correspondence continues to the third order with due presence of the axion pressure term in the small-scale: for the relativistic continuity and Euler equations of axion, see Eqs.\ (\ref{continuity-eq}) and (\ref{Euler-eq}). We show that the axion pressure term classically derived in Einstein's gravity (to third order) coincides exactly with the one derived from non-relativistic quantum mechanics in the literature. We show that from the third order the axion generates the rotational perturbation: see Eq.\ (\ref{axion-rotation}). For extremely low mass axion where the axion pressure term becomes cosmologically important, we note that the third order pure general relativistic correction terms are also affected by the axion pressure term: see Eqs.\ (\ref{varphi-dot-eq}) and (\ref{varphi-axion}).

Readers who want to skip the technical details of deriving the general relativistic continuity and Euler equation for axion fluid based on relativistic nonlinear perturbation analysis can go directly to Secs.\ \ref{sec:correspondence} and \ref{sec:discussion} after reading first two paragraphs of Sec.\ \ref{sec:axion}. For the benefit of such readers, below we briefly explain the basic method of analysis in the case of axion where we need fluid-field mixed formulation of nonlinear cosmological perturbation with proper time averaging.

As the axion we consider a classical minimally coupled massive scalar field and take temporal averaging over the coherent oscillation. The Einstein's equation becomes
\bea
   & & \widetilde G_{ab} = 8 \pi G \langle \widetilde T_{ab} \rangle,
   \label{method}
\eea
where we take the temporal averaging for the scalar field to get the averaged energy-momentum tensor, and the averaged fluid (density, pressure and velocity) quantities; although the average should be applied to the field quantities, here only to show the method we impose it to the energy-momentum tensor.
For the energy conservation equation we can use
\bea
   & & \langle \widetilde T^b_{0;b} \rangle = 0, \quad {\rm or} \quad
       \langle \widetilde T^b_{a;b} \widetilde n^a \rangle = 0,
   \label{method-E-conservation}
\eea
where the latter expression is the ADM energy-conservation with $\widetilde n_a$ the normal-frame four-vector. We note that in the case of axion it is troublesome to use the covariant energy conservation $\langle \widetilde T^b_{a;b} \widetilde u^b \rangle = 0$ as the fluid four-vector $\widetilde u^b$ also involves the scalar field \cite{fNL-2013,Noh-2014}. For the massive scalar field $\widetilde T_{ab}$ involves the scalar field in quadratic combination only, see Eq.\ (\ref{Tab-MSF}). For the equation of motion we do not take average, see Eq.\ (\ref{EOM-axion}). As the axion-comoving gauge we impose the scalar-type (longitudinal) perturbation part of $\langle \widetilde T^0_i \rangle$ equals to zero to all orders in perturbation, see Eq.\ (\ref{axion-CG}). We will use the fully nonlinear equations expressed using the fluid quantities \cite{fNL-2013,Noh-2014}; one important point in the analysis of axion is to first express the fluid quantities in terms of the scalar field, and then take average over the quadratic combination of scalar field.

We set $c \equiv 1 \equiv \hbar$ except for Sec.\ \ref{sec:correspondence}.

%
%
%
\section{Metric and fluid quantities}

We consider scalar- and vector-type perturbations in the Friedmann world model with the general background curvature, $\overline K$. Our metric convention valid to the nonlinear order is \cite{Bardeen-1988} \bea
   d s^2
   &=& - a^2 \left( 1 + 2 \alpha \right) d \eta^2
       - 2 a^2 ( \beta_{,i} + B^{(v)}_i )
       d \eta d x^i
   \nonumber \\
   & &
       + a^2 \Big[ \left( 1 + 2 \varphi \right) \gamma_{ij}
       + 2 \gamma_{,i|j}
       + C^{(v)}_{i|j} + C^{(v)}_{j|i}
       + 2 C^{(t)}_{ij}
       \Big]
       d x^i d x^j,
\eea where $a$ is the cosmic scale factor; $\alpha$, $\beta$, $\gamma$, $\varphi$, $B_i^{(v)}$, $C_i^{(v)}$ and $C^{(t)}_{ij}$ are arbitrary functions of space and time with $B^{(v)i}_{\;\;\;\;\;\;|i} \equiv 0 \equiv C^{(v)i}_{\;\;\;\;\;\;|i}$ for the vector-type perturbations and $C^{(t)i}_{\;\;\;\;\;i} \equiv 0 \equiv C^{(t)j}_{\;\;\;\;\;i|j}$ for the tensor-type perturbation; the spatial indices of perturbation variables are raised and lowered by $\gamma_{ij}$ as the metric, and the vertical bar indicates the covariant derivative based on $\gamma_{ij}$ as the metric; for concrete forms of $\gamma_{ij}$, see \cite{Noh-2014}. The scalar-, vector- and tensor-types decomposition is generally valid considering fully nonlinear perturbations \cite{York-1973}. To the nonlinear order the three-types of perturbations couple with each other in the equation level. In this work we {\it ignore} the tensor-type perturbation $C^{(t)}_{ij}$.

The energy-momentum tensor in the energy-frame ($\widetilde q_a \equiv 0$) is given as
\bea
   & & \widetilde T_{ab} = \widetilde \mu \widetilde u_a \widetilde u_b
       + \widetilde p \left( \widetilde g_{ab}
       + \widetilde u_a \widetilde u_b \right)
       + \widetilde \pi_{ab},
   \label{Tab-fluid}
\eea where $\widetilde \mu$, $\widetilde p$, $\widetilde u_a$ and $\widetilde \pi_{ab}$ are the energy density, pressure, fluid four-vector, and the anisotropic stress, respectively \cite{Ehlers-Ellis}; we have $\widetilde u^c \widetilde u_c \equiv -1$, $\widetilde \pi_{ab} = \widetilde \pi_{ba}$ and $\widetilde \pi^b_b \equiv 0 \equiv \widetilde \pi^b_{a;b}$. We introduce the perturbed order fluid three-velocity $\widehat v_i$ as \cite{fNL-2013}
\bea
   & & \widetilde u_i
       \equiv a \widehat \gamma \widehat v_i, \quad
       \widehat \gamma \equiv {1 \over \sqrt{ 1 -
       {\widehat v^k \widehat v_k \over 1 + 2 \varphi}}}.
   \label{u_i}
\eea
The index of $\widehat v_i$ is raised and lowered by $\gamma_{ij}$ as the metric; we ignore the anisotropic stress $\widetilde \pi_{ab}$ as the minimally coupled scalar field does not support the anisotropic stress.

As the spatial gauge condition we take
\bea
   & & \gamma \equiv 0 \equiv C^{(v)}_i,
   \label{spatial-gauge}
\eea
on the metric tensor.
This is the only spatial gauge condition which (together with our temporal gauge condition) allows the remaining variables to be gauge invariant to the fully nonlinear order \cite{Bardeen-1988,Noh-Hwang-2004,fNL-2013}. With this spatial gauge condition we set $\chi_i \equiv a (\beta_{,i} + B_i^{(v)} ) \equiv \chi_{,i} + \chi_i^{(v)}$, and the metric becomes
\bea
   & & d s^2 = - a^2 \left( 1 + 2 \alpha \right) d \eta^2
       - 2 a \chi_i d \eta d x^i
       + a^2 \left( 1 + 2 \varphi \right) \gamma_{ij}
       d x^i d x^j.
   \label{metric-FNL}
\eea
This allows us to expand the cosmological perturbations to fully nonlinear order in an exact manner \cite{fNL-2013,Noh-2014}.

\section{Minimally Coupled Scalar Field}

A minimally coupled scalar field is given as
\bea
   & & \widetilde T_{ab}
       = \widetilde \phi_{,a} \widetilde \phi_{,b}
       - \left[ {1 \over 2} \widetilde \phi^{;c} \widetilde \phi_{,c}
       + \widetilde V (\widetilde \phi) \right] \widetilde g_{ab}.
   \label{Tab-MSF}
\eea
In the energy-frame we can show $\widetilde \pi_{ab} = 0$ \cite{fNL-2013,Noh-2014}.

The fluid quantities follow from Eqs.\ (\ref{Tab-fluid}) and (\ref{Tab-MSF})
[see Eq.\ (5.6) of \cite{Noh-2014}]
\bea
   & & \widetilde \mu = {1 \over 2} {\widetilde {\dot {\widetilde \phi}}}^2
       + \widetilde V, \quad
       \widetilde p = {1 \over 2} {\widetilde {\dot {\widetilde \phi}}}^2
       - \widetilde V, \quad
       \widehat v_i = - {\widetilde \phi_{,i}
       \over a \widehat \gamma {\widetilde {\dot {\widetilde \phi}}}},
   \label{fluids-MSF}
\eea
where
\bea
   & & {\widetilde {\dot {\widetilde \phi}}}
       \equiv \widetilde \phi_{,c} \widetilde u^c
       = {1 \over \widehat \gamma} {D \widetilde \phi \over D t}, \quad
       {D \widetilde \phi \over Dt}
       \equiv {1 \over {\cal N}}
       \left( {\partial \over \partial t}
       + {\chi^i \over a^2 (1 + 2 \varphi)} \nabla_i \right)\widetilde\phi,
   \nonumber \\
   & &
       \widehat \gamma
       \equiv {1 \over \sqrt{ 1 - {\widetilde \phi^{|i} \widetilde \phi_{,i}
       \over a^2 ( 1 + 2 \varphi ) (D \widetilde \phi/Dt)^2}}}, \quad
       {\cal N} \equiv \sqrt{
       1 + 2 \alpha + {\chi^k \chi_k \over a^2 (1 + 2 \varphi)}}.
\eea

The equation of motion, $\widetilde \phi^{;c}_{\;\;\;c} = \partial \widetilde V / (\partial \widetilde \phi)$, gives [see Eq.\ (5.10) of \cite{Noh-2014}]
\bea
   \fl \ddot {\widetilde \phi}
       + \left( 3 H {\cal N}
       - {\cal N} \kappa
       - {\dot {\cal N} \over {\cal N}}
       - {\chi^i {\cal N}_{,i} \over
       a^2 {\cal N} ( 1 + 2 \varphi )} \right) \dot {\widetilde \phi}
       + {\cal N}^2 {\partial \widetilde V \over \partial \widetilde \phi}
       + {2 \chi^i \over a^2 ( 1 + 2 \varphi )} \dot {\widetilde \phi}_{,i}
   \nonumber \\
   \fl \qquad
       - {1 \over a^2 (1 + 2 \varphi)}
       \left( {\cal N}^2 \gamma^{ij}
       - {\chi^i \chi^j \over a^2 (1 + 2 \varphi)} \right) \widetilde \phi_{,i|j}
       + \Bigg[ - {{\cal N}^2 \over a^2 (1 + 2 \varphi)}
       \left( {{\cal N}^{|i} \over {\cal N}}
       + {\varphi^{|i} \over 1 + 2 \varphi} \right)
   \nonumber \\
   \fl \qquad
       + \left( 3 H {\cal N}
       - {\cal N} \kappa
       - {\dot {\cal N} \over {\cal N}}
       - {\chi^k {\cal N}_{,k} \over
       a^2 {\cal N} ( 1 + 2 \varphi )} \right)
       {\chi^i \over a^2 (1 + 2 \varphi)}
   \nonumber \\
   \fl \qquad
       + \left( {\chi^i \over a^2 (1 + 2 \varphi)}
       \right)^{\displaystyle\cdot}
       + {\chi^k \over a^4 (1 + 2 \varphi)}
       \left( {\chi^i \over 1 + 2 \varphi} \right)_{|k}
       \Bigg] \widetilde \phi_{,i}
       = 0.
   \label{EOM}
\eea

\section{Axion}
                                  \label{sec:axion}

We consider the axion as a massive scalar field with a potential
$ \widetilde V (\widetilde \phi) = {1 \over 2} m^2 \widetilde \phi^2$. The pseudo nature of axion is not relevant in our cosmological consideration. We consider only the classical nature of axion; we are not capable of participating in the ongoing controversies about the potential roles of quantum nature of the axion as a Bose-Einstein condensation \cite{Sikivie-Yang-2009,BEC-controversy}; see below Eq.\ (\ref{axion-pressure}) though.
We have \bea
   & & {H \over m} = 2.133 \times 10^{-28} h
       \left( { 10^{-5} {\rm eV} \over m } \right)
       \left( {H \over H_0} \right),
   \label{Hm}
\eea where $H_0 \equiv 100h \, {\rm km}{\rm sec}^{-1} {\rm Mpc}^{-1}$ is the present Hubble parameter with $H \equiv \dot a/a$. As the axion coherently oscillates we strictly {\it ignore} ${H / m}$ higher order terms.

We decompose the field and fluid quantities to the background and perturbed parts as
\bea
   & & \widetilde \phi \equiv \phi + \delta \phi, \quad
       \widetilde \mu \equiv \mu + \delta \mu, \quad
       \widetilde p \equiv p + \delta p.
\eea
We have a background solution \cite{Ratra-1991,Axion-1997}
\bea
   \phi (t) = a^{-3/2} \left[ \phi_{+0} \sin{(mt)}
       + \phi_{-0} \cos{(mt)} \right],
   \label{BG-phi}
\eea
where $\phi_{+0}$ and $\phi_{-0}$ are the constant coefficients. We take average over time scale of order $m^{-1}$ for all fluid quantities associated with the axion as a oscillating scalar field. \cite{Ratra-1991}. We have \cite{Ratra-1991,Axion-1997} \bea
   & & \mu
       = {1 \over 2} m^2 a^{-3} \left( \phi_{+0}^2 + \phi_{-0}^2 \right), \quad
       p = 0.
\eea
Thus, for the background the axion evolves exactly the same as a pressureless ideal fluid \cite{Axion-CDM}. This conclusion is valid in the presence of both the spatial curvature and the cosmological constant $\Lambda$ in the background, and in the presence of other fluids. For the nonlinear perturbation, here we consider a single presence of the axion field; considering the presence of other fluids and fields is trivial and those are not affected by the special nature of axion fluid.

\subsection{Fluid quantities and equation of motion}

Including perturbations we expand \cite{Ratra-1991,Axion-1997}
\bea
   & & \widetilde \phi \equiv \phi + \delta \phi
       \equiv ( a^{-3/2} \phi_{+0} + \delta \phi_+ ) \sin{(mt)}
       + ( a^{-3/2} \phi_{-0} + \delta \phi_- ) \cos{(mt)},
   \label{ansatz}
\eea
to fully nonlinear order, where $\delta \phi_+$ and $\delta \phi_-$ are arbitrary functions of space and time. In the analysis we ignore $H/m$ higher order terms with $\delta \dot \phi_{\pm} \sim H \delta \phi_{\pm}$.

For the fluid quantities, from Eq.\ (\ref{fluids-MSF}), to the fully nonlinear order, we can show
\bea
   \fl \Bigg\{
            \begin{array}{c}
                \widetilde \mu \\
                \widetilde p
            \end{array}
       \Bigg\}
       = {1 \over 2} \langle {\widetilde {\dot {\widetilde \phi}}}^2 \rangle
       \pm {1 \over 2} m^2 \langle \widetilde \phi^2 \rangle
   \nonumber \\
   \fl \qquad
       = {1 \over 2 {\cal N}^2}
       \left[
       \langle \dot {\widetilde \phi}^2 \rangle
       + 2 {\chi^i \over a^2 ( 1 + 2 \varphi )}
       \langle \dot {\widetilde \phi} \widetilde \phi_{,i} \rangle
       + {\chi^i \chi^j \over a^4 (1 + 2 \varphi)^2}
       \langle \widetilde \phi_{,i} \widetilde \phi_{,j} \rangle
       \right]
   \nonumber \\
   \fl \qquad \qquad
       - {1 \over 2 a^2 (1 + 2 \varphi)}
       \langle \widetilde \phi^{,i} \widetilde \phi_{,i} \rangle
       \pm {1 \over 2} m^2 \langle \widetilde \phi^2 \rangle
   \nonumber \\
   \fl \qquad
       = m^2 {1 \over 4} \left( {1 \over {\cal N}^2} \pm 1 \right)
       \left[ a^{-3} \left( \phi_{+0}^2 + \phi_{-0}^2 \right)
       + 2 a^{-3/2} \left( \phi_{+0} \delta \phi_+
       + \phi_{-0} \delta \phi_- \right)
       + \delta \phi_+^2 + \delta \phi_-^2 \right]
   \nonumber \\
   \fl \qquad \qquad
       + m {\chi^i \over 2 a^2 {\cal N}^2 (1 + 2 \varphi)}
       \left[
       a^{-3/2} \left( \phi_{+0} \delta \phi_{-,i}
       - \phi_{-0} \delta \phi_{+,i} \right)
       + \delta \phi_+ \delta \phi_{-,i}
       - \delta \phi_- \delta \phi_{+,i} \right]
   \nonumber \\
   \fl \qquad \qquad
       + {\chi^i \chi^j \over 4 a^4 {\cal N}^2 ( 1 + 2 \varphi )^2}
       \left( \delta \phi_{+,i} \delta \phi_{+,j}
       + \delta \phi_{-,i} \delta \phi_{-,j} \right)
   \nonumber \\
   \fl \qquad \qquad
       - {1 \over 4 a^2 ( 1 + 2 \varphi )}
       \left( \delta \phi_+^{\;\;|i} \delta \phi_{+,i}
       + \delta \phi_-^{\;\;|i} \delta \phi_{-,i} \right),
   \label{mu-p-axion}
\eea
where the upper and lower signs correspond to the $\widetilde \mu$ and $\widetilde p$ parts, respectively. In the temporal comoving gauge (hypersurface) condition we will take, the $\widehat v_i$ part is fixed by the gauge condition, and later we will identify other variable as the velocity perturbation, see Eqs.\ (\ref{kappa-def}) and (\ref{kappa-u}). In the axion case where we need proper averaging, the parameter $\widehat v_i$ becomes ambiguous, see paragraph containing Eq.\ (\ref{axion-CG}).

The equation of motion in Eq.\ (\ref{EOM}), to the fully nonlinear order, gives
\bea
   \fl \Bigg\{ a^{-3/2} \left( {\cal N} - 1 \right)
       \left[ \mp 3 H m \phi_{\mp0} + m^2 \left( 1 + {\cal N} \right) \phi_{\pm0} \right]
   \nonumber \\
   \fl \qquad
       \mp 2 m \delta \dot \phi_\mp \mp 3 H m {\cal N} \delta \phi_\mp
       + \delta \ddot \phi_\pm
       + 3 H {\cal N} \delta \dot \phi_\pm
       + m^2 \left( {\cal N}^2 - 1 \right) \delta \phi_\pm
   \nonumber \\
   \fl \qquad
       - \left( {\cal N} \kappa
       + {\dot {\cal N} \over {\cal N}}
       + {\chi^i {\cal N}_{,i} \over
       a^2 {\cal N} ( 1 + 2 \varphi )} \right)
       \left[ a^{-3/2} \left( \mp m \phi_{\mp0}
       - {3 \over 2} H \phi_{\pm0} \right)
       \mp m \delta \phi_\mp + \delta \dot \phi_\pm \right]
   \nonumber \\
   \fl \qquad
       + {2 \chi^i \over a^2 ( 1 + 2 \varphi )}
       \left( \mp m \delta \phi_{\mp,i}
       + \delta \dot \phi_{\pm,i} \right)
       - {1 \over a^2 (1 + 2 \varphi)}
       \left( {\cal N}^2 \gamma^{ij}
       - {\chi^i \chi^j \over a^2 (1 + 2 \varphi)} \right)
       \delta \phi_{\pm,i|j}
   \nonumber \\
   \fl \qquad
       + \Bigg[ - {{\cal N}^2 \over a^2 (1 + 2 \varphi)}
       \left( {{\cal N}^{|i} \over {\cal N}}
       + {\varphi^{|i} \over 1 + 2 \varphi} \right)
   \nonumber \\
   \fl \qquad
       + \left( 3 H {\cal N}
       - {\cal N} \kappa
       - {\dot {\cal N} \over {\cal N}}
       - {\chi^k {\cal N}_{,k} \over
       a^2 {\cal N} ( 1 + 2 \varphi )} \right)
       {\chi^i \over a^2 (1 + 2 \varphi)}
   \nonumber \\
   \fl \qquad
       + \left( {\chi^i \over a^2 (1 + 2 \varphi)}
       \right)^{\displaystyle\cdot}
       + {\chi^k \over a^4 (1 + 2 \varphi)}
       \left( {\chi^i \over 1 + 2 \varphi} \right)_{|k}
       \Bigg]
       \delta \phi_{\pm,i}
       \Bigg\}
       \times
       \Bigg\{
            \begin{array}{c}
                \sin{(mt)} \\
                \cos{(mt)}
            \end{array}
       \Bigg\}
       = 0,
   \label{EOM-axion}
\eea
where the upper and lower signs correspond to the $\sin$ and $\cos$ parts, respectively; here we have not ignored $H/m$-order terms yet.

\subsection{Axion-comoving gauge}

We can decompose
\bea
   & &
       \widehat v_i \equiv - \widehat v_{,i} + \widehat v_i^{(v)}, \quad
       \widehat v^{(v)i}_{\;\;\;\;\;\;|i} \equiv 0.
   \label{v-decomposition}
\eea
In the fluid case the comoving gauge takes
\bea
   & & \widehat v \equiv 0,
\eea
as the temporal hypersurface (slicing) condition. For a scalar field the comoving gauge is the same as $\delta \phi \equiv 0$, the uniform-field gauge to fully nonlinear order \cite{fNL-2013,Noh-2014}.

In the axion case, however, as we have to take proper average over the oscillating field, the concept of $\widehat v_i$ and its decomposition become ambiguous. Thus, for the temporal gauge condition we go back to the original energy-momentum tensor and take the scalar part of $\widetilde T^0_i$ equal to zero as the comoving gauge. Thus, in the axion-comoving gauge we can impose
\bea
   & & \langle \widetilde T^0_i \rangle^{|i} \equiv 0,
   \label{axion-CG}
\eea
to all perturbation orders. Together with the spatial gauge condition in Eq.\ (\ref{spatial-gauge}) the axion-comoving slicing condition completely fixes the gauge mode. Each perturbation variable in these conditions has a unique gauge-invariant counterpart to all orders in perturbation \cite{Bardeen-1988,Noh-Hwang-2004,fNL-2013}.

Using the inverse metric presented in \cite{Noh-2014}, from Eq.\ (\ref{Tab-MSF}) we can show
\bea
   \fl \langle \widetilde T^0_i \rangle
       = - {1 \over a {\cal N}^2}
       \left( \langle \dot {\widetilde \phi} \widetilde \phi_{,i} \rangle
       + {\chi^j \over
       a^2 ( 1 + 2 \varphi )} \langle \widetilde \phi_{,i} \widetilde \phi_{,j} \rangle \right)
   \nonumber \\
   \fl \qquad
       = - {1 \over 2 a {\cal N}^2} \Bigg[
       a^{-3/2} m
       \left( \phi_{+0} \delta \phi_{-,i}
       - \phi_{-0} \delta \phi_{+,i} \right)
       + m \left( \delta \phi_{+} \delta \phi_{-,i}
       - \delta \phi_{-} \delta \phi_{+,i} \right)
   \nonumber \\
   \fl \qquad \qquad
       + {1 \over a^2 (1 + 2 \varphi)}
       \left( \delta \phi_{+,i} \delta \phi_{+,j}
       + \delta \phi_{-,i} \delta \phi_{-,j} \right) \chi^j \Bigg].
   \label{T_0i}
\eea
As we mentioned below Eq.\ (\ref{method}) we can omit average notation for $\widetilde T_{ab}$. Up to this point all equations are valid to fully nonlinear order.

Now, we consider perturbations to the third order. From the gauge condition in Eq.\ (\ref{axion-CG}) we can show
\bea
   & & {\delta \phi_+ \over \phi_{+0}}
       - {\delta \phi_- \over \phi_{-0}}
       = {a \over m} a^{-3/2}
       {\phi_{+0}^2 + \phi_{-0}^2 \over \phi_{+0}^3 \phi_{-0}}
       \Delta^{-1} \nabla^i \left(
       \delta \phi_{+,i} \delta \phi_{+,j} \chi^j
       \right),
\eea
to the third order, Thus
\bea
   & & \widetilde T^0_i
       = - {\mu \over m^2 \phi_{+0}^2}
       \left[ \delta \phi_{+,i} \delta \phi_{+,j} \chi^j
       - \nabla_i \Delta^{-1} \nabla^j \left(
       \delta \phi_{+,j} \delta \phi_{+,k} \chi^k
       \right) \right],
\eea
which is pure third-order vector-type perturbation. In the fluid formulation
Eqs.\ (\ref{Tab-fluid}) and (\ref{u_i}) give
\bea
   & & \widetilde T^0_i = \left( \widetilde \mu + \widetilde p \right)
       {\widehat \gamma^2 \over {\cal N}} \widehat v_i.
\eea
By introducing $\widehat v_i$ in this fluid relation and using Eq.\ (\ref{v-decomposition}), we have
\bea
   & & \widehat v = 0,
   \nonumber \\
   & &
       \widehat v_i^{(v)}
       = {1 \over \mu} \widetilde T^0_i
       = - {1 \over m^2 \phi_{+0}^2}
       \left[ \delta \phi_{+,i} \delta \phi_{+,j} \chi^j
       - \nabla_i \Delta^{-1} \nabla^j \left(
       \delta \phi_{+,j} \delta \phi_{+,k} \chi^k
       \right) \right].
   \label{v-vector}
\eea
Thus, axion has the rotational perturbation appearing from the third order; from Eq.\ (\ref{u_i}) we have $\widetilde u_i = a \widehat v_i^{(v)}$.

To the third order in the axion comoving gauge Eq.\ (\ref{mu-p-axion}) can be arranged as
\bea
   & & \Bigg\{
            \begin{array}{c}
                \widetilde \mu \\
                \widetilde p
            \end{array}
       \Bigg\}
       = \mu {1 \pm {\cal N}^2 \over 2 {\cal N}^2}
       \left[ 1 + 2 {a^{3/2} \over \phi_{+0}^2 + \phi_{-0}^2}
       \left( \phi_{+0} \delta \phi_+
       + \phi_{-0} \delta \phi_- \right)
       + {a^3 \over \phi_{+0}^2}
       \delta \phi_+^2 \right]
   \nonumber \\
   & & \qquad
       - \mu {1 \over 2 m^2 a^2 ( 1 + 2 \varphi )}
       {a^3 \over \phi_{+0}^2} \delta \phi_+^{\;\;|i} \delta \phi_{+,i}.
   \label{mu-p-axion-third}
\eea
These are the energy density and pressure supported by the axion fluid to the third-order perturbation. Besides the Einstein's equations to be presented below, we need an expression of $\delta {\cal N}$ in terms of the axion field. This relation follows from the equation of motion, and will be presented in Eq.\ (\ref{delta-N}).

\subsection{Einstein equations}

Equations valid to fully nonlinear perturbation orders are presented in \cite{fNL-2013,Noh-2014}. Here we will again borrow some equations from these works.

As explained in the paragraph containing Eq.\ (\ref{method}) we can use the equations expressed in terms of fluid quantities but have to change fluid quantities to the scalar field before we take average over the oscillating field. After averaging, we will strictly ignore $H/m$-order terms, and will keep only first order in $\Delta/(m^2 a^2)$. It is convenient to have
\bea
   & & \kappa = \dot \delta \sim H \delta, \quad
       \chi_i = \chi_{,i} = - {a^2 \over \Delta} \kappa_{,i},
\eea
to the linear order.

From the covariant energy conservation and the trace of ADM (Arnowitt-Deser-Misner) propagation equations in Eqs.\ (3.8) and (3.4) of \cite{Noh-2014}, respectively, to the third order, we have
\bea
   & &
       \dot {\widetilde \mu}
       + {1 \over a^2 (1 + 2 \varphi)} \delta \mu_{,k} \chi^k
       + {\cal N} \left( \widetilde \mu + \widetilde p \right)
       \left( 3 H - \kappa \right)
       = 0,
   \label{energy-conservation} \\
   & &
       - 3 \left[ {1 \over {\cal N}}
       \left( {\dot a \over a} \right)^{\displaystyle\cdot}
       + {\dot a^2 \over a^2}
        + {4 \pi G \over 3} \left( \widetilde \mu + 3 \widetilde p \right)
       - {\Lambda \over 3} \right]
       + {1 \over {\cal N}} \dot \kappa
       + 2 {\dot a \over a} \kappa
       + {\Delta {\cal N} \over a^2 {\cal N} (1 + 2 \varphi)}
   \nonumber \\
   & & \qquad
       = {1 \over 3} \kappa^2
       - {1 \over a^2 {\cal N} (1 + 2 \varphi)} \left(
       \chi^{i} \kappa_{,i}
       + {\varphi^{|i} {\cal N}_{,i} \over 1 + 2 \varphi} \right)
       + \overline{K}^i_j \overline{K}^j_i,
   \label{Raychaudhury-eq}
\eea
where from Eq.\ (3.10) in \cite{Noh-2014},
\bea
   & & \overline{K}^i_j \overline{K}^j_i
       = {1 \over a^4 {\cal N}^2 (1 + 2 \varphi)^2}
       \Bigg\{
       {1 \over 2} \chi^{i|j} \left( \chi_{i|j} + \chi_{j|i} \right)
       - {1 \over 3} \chi^i_{\;\;|i} \chi^j_{\;\;|j}
   \nonumber \\
   & & \qquad
       - 4 \left[
       {1 \over 2} \chi^i \varphi^{|j} \left(
       \chi_{i|j} + \chi_{j|i} \right)
       - {1 \over 3} \chi^i_{\;\;|i} \chi^j \varphi_{,j} \right] \Bigg\}.
   \label{K-bar-eq}
\eea
to the third order. A perturbation variable $\kappa$ is defined as
\bea
   & & K^i_i \equiv - 3 {\dot a \over a} + \kappa
       \equiv - \widetilde \theta^{(n)} \equiv - \widetilde n^c_{\;\; ;c},
   \label{kappa-def}
\eea where $K_{ij}$ is the extrinsic curvature, and $\widetilde \theta^{(n)}$ is the expansion scalar of the normal-frame four-vector; later we will identify $\kappa$ as the divergence of velocity perturbation, see Eq.\ (\ref{kappa-u}).
We also need the ADM energy constraint and ADM momentum constraint equations to the second order. From Eqs.\ (3.2) and (3.3) in \cite{Noh-2014}, respectively, we have
\bea
   \fl {\Delta \over a^2} \varphi
       = - 4 \pi G \varrho \delta - {\dot a \over a} \kappa
       + {1 \over 6} \kappa^2
       + {1 \over a^2} \left( 4 \varphi \Delta \varphi
       + {3 \over 2} \varphi^{|i} \varphi_{,i} \right)
       - 3 \varphi \left( 1 - 2 \varphi \right)
       {\overline K \over a^2}
   \nonumber \\
   \fl \qquad
       - {1 \over 4 a^4} \left[ {1 \over 2} \chi^{i|j}
       \left( \chi_{i|j} + \chi_{j|i} \right)
       - {1 \over 3} \chi^i_{\;\;|i} \chi^j_{\;\;|j} \right],
   \label{E-constraint} \\
   \fl {2 \over 3} \kappa_{,i}
       + {1 \over a^2 {\cal N} ( 1 + 2 \varphi )}
       \left[ {1 \over 2} \left( \Delta \chi_i
       + \chi^k_{\;\;|ik} \right)
       - {1 \over 3} \chi^k_{\;\;|ki} \right]
   \nonumber \\
   \fl \qquad
       =
       {1 \over a^2}
       \Bigg\{
       \left( {\cal N} - \varphi \right)_{,j}
       \left[ {1 \over 2} \left( \chi^{j}_{\;\;|i} + \chi_i^{\;|j} \right)
       - {1 \over 3} \delta^j_i \chi^k_{\;\;|k} \right]
       + \nabla_j
       \left(
       \chi^{j} \varphi_{,i}
       + \chi_{i} \varphi^{|j}
       - {2 \over 3} \delta^j_i \chi^{k} \varphi_{,k} \right)
       \Bigg\},
   \label{mom-constraint}
\eea
valid to the second order.


\subsection{Results}

Now we {\it assume} a flat background curvature, $\overline K \equiv 0$.
From the equation of motion in Eq.\ (\ref{EOM-axion}) we can show
\bea
   & & \delta {\cal N}
       = \left( 1 - \Phi + \Phi^2 \right) {\Delta \over 2 m^2 a^2} \Phi,
   \label{delta-N}
\eea
where $\Phi \equiv {\delta \phi_+ /( a^{-3/2} \phi_{+0})}$, and we set ${\cal N} \equiv 1 + \delta {\cal N}$. This relation follows from dividing Eq.\ (\ref{EOM-axion}) by $\phi_{\mp 0}$ and summing the two relations.

From Eq.\ (\ref{mu-p-axion-third}) we have
\bea
   & & \delta = 2 \Phi + \Phi^2
       - \left( 1 + \Phi \right) {\Delta \over 2 m^2 a^2} \Phi
       - {1 \over 2 m^2 a^2} \Phi^{,i} \Phi_{,i},
   \nonumber \\
   & &
       {\delta p \over \mu}
       = - \left( 1 + \Phi \right)
       {\Delta \over 2 m^2 a^2} \Phi
       - {1 \over 2 m^2 a^2} \Phi^{,i} \Phi_{,i},
   \nonumber \\
   & &
       \Phi = {1 \over 2} \left( 1 + {\Delta \over 4 m^2 a^2} \right)
       \left( \delta
       - {1 \over 4} \delta^2
       + {1 \over 8} \delta^3 \right)
       + {1 \over 16 m^2 a^2} \left( 1 - {3 \over 2} \delta \right)
       \delta^{,i} \delta_{,i}.
   \label{Phi-relation}
\eea
Thus, we have
\bea
   \fl \delta {\cal N}
       = {1 \over 4 m^2} \Bigg[
       \left( 1 - {1 \over 2} \delta + {3 \over 8} \delta^2 \right)
       {\Delta \over a^2} \delta
       - {1 \over 4} \left( 1 - {1 \over 2} \delta \right)
       {\Delta \over a^2} \delta^2
       + {1 \over 8} {\Delta \over a^2} \delta^3 \Bigg]
       = {1 \over 2 m^2 a^2} {\Delta \sqrt{1 + \delta}
       \over \sqrt{1 + \delta}},
   \nonumber \\
   \fl {\delta p \over \mu}
       = {1 \over 4 m^2} \Bigg[
       - \left( 1 + {1 \over 2} \delta - {1 \over 8} \delta^2 \right)
       {\Delta \over a^2} \delta
       + {1 \over 4} \left( 1 + {1 \over 2} \delta \right)
       {\Delta \over a^2} \delta^2
       - {1 \over 8} {\Delta \over a^2} \delta^3
       - {1 \over 2 a^2} \left( 1 - \delta \right)
       \delta^{,i} \delta_{,i} \Bigg],
   \label{delta-N-p}
\eea
where the compact expression in $\delta {\cal N}$ is motivated by the known expression in the literature on non-relativistic Bose-Einstein condensation in Minkowski space-time \cite{BEC}, see below Eq.\ (\ref{axion-pressure}). The ADM momentum conservation in Eq.\ (3.7) of \cite{Noh-2014} gives
\bea
   & & \widetilde p_{,i} + \left( \widetilde \mu + \widetilde p \right)
       {{\cal N}_{,i} \over {\cal N}}
       = - \mu {1 \over m^2 a^2} \left( \Phi^{,j} \Phi_{,i} \right)_{,j},
   \label{ADM-mom-conservation}
\eea
and this is consistent with Eqs.\ (\ref{delta-N}) and (\ref{Phi-relation}).

Now, Eqs.\ (\ref{energy-conservation}), (\ref{Raychaudhury-eq}) and (\ref{mom-constraint}), respectively, give
\bea
   \fl \dot \delta - \kappa
       + {1 \over a^2} \left( 1 - 2 \varphi \right)
       \delta_{,i} \chi^i
       - \delta \kappa
       = 0,
   \label{energy-conservation-correction} \\
   \fl \dot \kappa
       + 2 H \kappa
       - 4 \pi G \mu \delta
       + {1 \over a^2} \left( 1 - 2 \varphi \right) \kappa_{,i} \chi^i
       - {1 \over 3} \kappa^2
   \nonumber \\
   \fl \qquad
       - {1 \over a^4} \left( 1 - 4 \varphi \right)
       \left[ {1 \over 2} \chi^{i,j} \left( \chi_{i,j} + \chi_{j,i} \right)
       - {1 \over 3} \left( \Delta \chi \right)^2 \right]
       + {4 \over a^4} \left(
       \chi^{,i} \varphi^{,j} \chi_{,ij}
       - {1 \over 3} \chi^{,i} \varphi_{,i} \Delta \chi \right)
   \nonumber \\
   \fl \qquad
       = - {\Delta \over a^2} \delta {\cal N}
       = - {\Delta \over 2 m^2 a^4} {\Delta \sqrt{1 + \delta}
       \over \sqrt{1 + \delta}},
   \label{Raychaudhury-eq-correction}
\eea
valid to the third order, and
\bea
   \fl \Big( \kappa + {\Delta \over a^2} \chi \Big)_{,i}
       + {3 \over 4} {\Delta \over a^2} \chi_i^{(v)}
       = {1 \over a^2}
       \Big[
       \left( 2 \varphi \Delta \chi
       - \varphi_{,j} \chi^{,j} \right)_{,i}
       + {3 \over 2} \left(
       \varphi_{,ij} \chi^{,j}
       + \chi_{,i} \Delta \varphi \right)
       \Big],
   \label{mom-constraint-correction}
\eea
valid to the second order.
We note that compared with the CDM case in \cite{Third-2005} the only correction from the axion nature appears in the right-hand-side of Eq.\ (\ref{Raychaudhury-eq-correction}).

From Eq.\ (\ref{mom-constraint-correction}) we have $\chi_i^{(v)} = 0$ to the linear order. To the second order
Eq.\ (\ref{mom-constraint-correction}) can be decomposed as
\bea
   & & \kappa + {\Delta \over a^2} \chi
       = {1 \over a^2} \Big[ 2 \varphi \Delta \chi
       - \varphi_{,i} \chi^{,i}
       + {3 \over 2} \Delta^{-1} \nabla^i \left(
       \varphi_{,ij} \chi^{,j}
       + \chi_{,i} \Delta \varphi \right) \Big]
       \equiv {1 \over a} X,
   \nonumber \\
   & & {\Delta \over a^2} \chi_i^{(v)}
       = {2 \over a^2} \Big[
       \varphi_{,ij} \chi^{,j}
       + \chi_{,i} \Delta \varphi
       - \nabla_i \Delta^{-1} \nabla^j \left(
       \varphi_{,jk} \chi^{,k}
       + \chi_{,j} \Delta \varphi \right)
       \Big]
       \equiv {1 \over a} Y_i,
   \nonumber \\
   & & {\Delta \over a^2} \chi_i
       = - \kappa_{,i} + {1 \over a} X_{,i} + {1 \over a} Y_i.
   \label{chi}
\eea

\section{Relativistic continuity and Euler equations for Axion}
                                  \label{sec:correspondence}

Now we have complete relations for axion fluid valid to the third order perturbation in Einstein's gravity. We will arrange the equations so that we can readily compare with the Newtonian fluid equations.

We {\it identify} $\delta$ as the density contrast ${\delta \varrho / \varrho}$, and $\kappa$ a perturbed part of trace of extrinsic curvature in Eq.\ (\ref{kappa-def}) as the divergence of velocity perturbation \cite{Noh-Hwang-2004,Third-2005}
\bea
   & & \kappa \equiv - {1 \over a} \nabla \cdot {\bf u}
       \equiv - {\Delta \over a} u.
   \label{kappa-u}
\eea
Here we assume the flat background curvature but include the cosmological constant. From Eq.\ (\ref{chi}), to the second order, we have
\bea
   & & {c \over a} \vec{\chi}
       = {\bf u}
       + \Delta^{-1} \left( \nabla X + {\bf Y} \right),
   \\
   & & X
       \equiv 2 \varphi \nabla \cdot {\bf u}
       - {\bf u} \cdot \nabla \varphi
       + {3 \over 2} \Delta^{-1} \nabla \cdot \left(
       {\bf u} \cdot \nabla \nabla \varphi
       + {\bf u} \Delta \varphi \right),
   \nonumber \\
   & & {\bf Y}
       \equiv 2 \Big[ {\bf u} \cdot \nabla \nabla \varphi
       + {\bf u} \Delta \varphi
       - \nabla \Delta^{-1} \nabla \cdot \left(
       {\bf u} \cdot \nabla \nabla \varphi
       + {\bf u} \Delta \varphi \right) \Big],
   \label{X-Y-def}
\eea
with $\nabla \cdot {\bf Y} = 0$; $\chi^i$ and $\varphi$ are metric variables in Eq.\ (\ref{metric-FNL}), interpreted as the perturbed shear (of the normal four-vector flow) and perturbed spatial curvature, respectively \cite{fNL-2013,Noh-2014}.

Equations (\ref{energy-conservation-correction}) and (\ref{Raychaudhury-eq-correction}) give the general relativistic version of the continuity (mass-conservation) and Euler (momentum-conservation) equations valid to third order perturbations in the axion case. These are
\bea
   & & \dot \delta
       + {1 \over a} \nabla \cdot {\bf u}
       + {1 \over a} \nabla \cdot \left( \delta {\bf u} \right)
       = {1 \over a} \left( \nabla \delta \right) \cdot
       \left[ 2 \varphi {\bf u}
           - \Delta^{-1} \left( \nabla X + {\bf Y} \right) \right],
   \label{continuity-eq} \\
   & & {1 \over a} \nabla \cdot \Big( \dot {\bf u}
       + {\dot a \over a} {\bf u} \Big)
       + 4 \pi G \varrho \delta
       + {1 \over a^2} \nabla \cdot \left( {\bf u} \cdot \nabla {\bf u} \right)
       - {\hbar^2 \Delta \over 2 m^2 a^4} {\Delta \sqrt{1 + \delta}
       \over \sqrt{1 + \delta}}
   \nonumber \\
   & & \qquad
       = {1 \over a^2} \Bigg\{
       - {2 \over 3} \varphi {\bf u} \cdot \nabla
       \left( \nabla \cdot {\bf u} \right)
       + 4 \nabla \cdot \left[
       \varphi \left( {\bf u} \cdot \nabla {\bf u}
       - {1 \over 3} {\bf u} \nabla \cdot {\bf u} \right) \right]
   \nonumber \\
   & & \qquad
       + {2 \over 3} X \nabla \cdot {\bf u}
       + {\bf u} \cdot \left( \nabla X + {\bf Y} \right)
       - \Delta \left[ {\bf u} \cdot \Delta^{-1}
       \left( \nabla X + {\bf Y} \right) \right] \Bigg\}.
   \label{Euler-eq}
\eea
In order to close the equations we need $\varphi$ valid to the linear order. From Eq.\ (\ref{E-constraint}) we have
\bea
   & &  c^2{\Delta \over a^2} \varphi
       = - 4 \pi G \varrho \delta
       + {\dot a \over a} {1 \over a} \nabla \cdot {\bf u}.
   \label{varphi-lin-eq}
\eea
Now Eqs.\ (\ref{X-Y-def})-(\ref{varphi-lin-eq}) give closed differential equations for density and velocity perturbations, $\delta$ and ${\bf u}$. These are valid in the presence of the cosmological constant.

In order to derive the axion contribution we have strictly ignored
\bea
   & & {\hbar H \over c^2 m},
\eea
order terms, and have kept up to only first order in
\bea
   & & {\hbar^2 k^2 \over c^2 m^2 a^2}
       = 4.088 \times 10^{-43}
       \left( {10^{-5} {\rm eV} \over m} \right)^2
       \left( {1 {\rm kpc} \over a_0/k} \right)^2
       \left( {a_0 \over a} \right)^2.
\eea

In this section we have recovered $c$ and $\hbar$. The curvature perturbation variable $\varphi$ is dimensionless with $\varphi \sim \delta \Phi/c^2$ where $\delta \Phi$ is the perturbed gravitational potential. In the Newtonian limit with $c$-goes-to-infinity, the right-hand-sides of Eqs.\ (\ref{continuity-eq}) and (\ref{Euler-eq}) vanishes, thus recovering, except for the axion correction term, the well known Newtonian continuity and Euler equations valid to fully nonlinear order \cite{Peebles-1980}; notice that in Einstein's gravity these equations without $\varphi$ corrections are valid only to the second order perturbation.

The presence of Planck's constant $\hbar$ in the axion correction term, despite of our classical analysis, can be traced to the presence of the constant in the potential
\bea
   & & \widetilde V (\widetilde \phi)
       = {1 \over 2} \left( {m c^2 \over \hbar} \right)^2
       \widetilde \phi^2,
\eea
with a dimension $[\widetilde \phi] = (M/L)^{1/2}$.

The term involving $m$ in Eq.\ (\ref{Euler-eq}) is the pressure correction term arising from the axion nature of the system. The only difference of axion from the zero-pressure fluid (CDM) is represented by this axion pressure term. The left-hand-sides of Eq.\ (\ref{continuity-eq}) and (\ref{Euler-eq}), except for the axion pressure term, are identical to the Newtonian continuity and Euler equations, respectively \cite{Peebles-1980}. The right-hand-side are third-order terms; these are pure general relativistic corrections in the zero-pressure fluid \cite{Third-2005} and are numerically negligible in the current paradigm of concordance cosmology for zero-pressure fluid \cite{JGNH-2011}; the same is true for axion in the super-Jeans scales, see below.

The axion pressure contribution can be written as
\bea
   & & c^2 {\Delta \over a^2} \delta {\cal N}
       = {\hbar^2 \Delta \over 2 m^2 a^4} {\Delta \sqrt{1 + \delta}
       \over \sqrt{1 + \delta}}
       = {\hbar^2 \Delta \over 2 m^2 a^4} {\Delta \sqrt{\widetilde \varrho}
       \over \sqrt{\widetilde \varrho}},
   \label{axion-pressure}
\eea
where $\widetilde \mu \equiv \widetilde \varrho c^2 \equiv (1 + \delta) \varrho c^2$. Equations (\ref{delta-N}), (\ref{Phi-relation}) and (\ref{delta-N-p}) show that the perturbed part of lapse function $\delta {\cal N}$ is only loosely related to the perturbed pressure, see Eq.\ (\ref{ADM-mom-conservation}). We note that this so-called axion pressure term derived from general relativistic third order perturbation {\it coincides} (except for the scale factor) exactly with the one derived from Schr\"odinger equation in the non-relativistic limit in Minkowski space-time \cite{BEC,Sikivie-Yang-2009}. Although the authors of \cite{Sikivie-Yang-2009} emphasized that the result is due to Bose-Einstein condensate nature of the axion, we stress that our result is a purely classical one based on perturbation treatment of a massive scalar field in Einstein's gravity without assuming non-relativistic limit; the resulting coincidence proves that the classical axion is non-relativistic indeed.

Equations (\ref{continuity-eq}) and (\ref{Euler-eq}) can be combined to give
\bea
   & & \ddot \delta + 2 {\dot a \over a} \dot \delta
       - 4 \pi G \varrho \delta
       + {1 \over a^2} \left[ a \nabla \cdot \left( \delta {\bf u} \right)
       \right]^{\displaystyle\cdot}
       - {1 \over a^2} \nabla \cdot \left( {\bf u}
       \cdot \nabla {\bf u} \right)
       + {\hbar^2 \Delta \over 2 m^2 a^4} {\Delta \sqrt{1 + \delta}
       \over \sqrt{1 + \delta}}
   \nonumber \\
   & & \qquad
       = \; {\rm third \; order \; terms},
   \label{ddot-delta-eq}
\eea
where third order terms in the right-hand-side are pure general relativistic terms all involving linear order $\varphi$ term. Besides the axion pressure term, the left-hand-side is valid to fully nonlinear order in Newtonian theory (this is valid in the Einstein's theory to the second order only).

Comparing the gravity and pressure terms to the linear order we have the axion Jeans scale \cite{Khlopov-etal-1985,Nambu-Sasaki-1990,Sikivie-Yang-2009,Axion-2009} \bea
   & & \lambda_J = {2 \pi a \over k_J}
       = \left( {\pi^3 \hbar^2 \over G \varrho m^2} \right)^{1/4}
       = 5.4 \times 10^{14} {\rm cm} h^{-1/2}
       \left( {m \over 10^{-5} {\rm eV}} \right)^{-1/2},
   \label{Jeans-scale}
\eea
with $\Delta = - k^2$.
Examination of Eq.\ (\ref{ddot-delta-eq}) reveals that the same criteria applies to the third order. The axion pressure term has a role on scale smaller than the Solar System size for a canonical mass $m \sim 10^{-5} {\rm eV}$, thus completely negligible in all cosmological scales. This proves the CDM nature of the axion in the super-Jeans scale.

Notice that the third order terms arising from the pure Einstein's gravity in the right-hand-sides of Eqs.\ (\ref{continuity-eq}), (\ref{Euler-eq}) and (\ref{ddot-delta-eq}) include linear order $\varphi$ term.
From Eqs.\ (\ref{continuity-eq})-(\ref{varphi-lin-eq}) we have
\bea
   & & \dot \varphi
       = {\dot a \over a} {\hbar^2 \Delta \over 4 c^2 m^2 a^2} \delta.
   \label{varphi-dot-eq}
\eea
to the linear order.
The homogeneous solution (which is constant in time) is the zero-pressure fluid part and the inhomogeneous solution is the additional contribution from the axion. From Eqs.\ (\ref{varphi-lin-eq}) and (\ref{varphi-dot-eq}) we have the fluid part and the additional axion part of $\varphi$ as
\bea
   & & \varphi |_{\rm fluid} \sim {a^2 H^2 \over c^2 \Delta} \delta, \quad
       \varphi |_{\rm axion} \sim {\hbar^2 \Delta \over 4 c^2 m^2 a^2} \delta.
   \label{varphi-axion}
\eea
Thus, the pure axion part of $\varphi$ is negligible in the super-Jeans scale. We have $\varphi |_{\rm fluid} = C({\bf x}) = {\rm constant}$ in time even in the presence of the cosmological constant, see Eq.\ (\ref{varphi-dot-eq}); $\varphi$ is the curvature perturbation in the comoving gauge
which is the well known conserved quantity in the super-sound-horizon scale \cite{HN-GGT-2005}. Even in sub-Jeans scale the pure axion part of $\varphi$ is simply decaying with $\varphi |_{\rm axion} \propto a^{-1}$ for $\Lambda = 0$.

Therefore, in the cosmological scales we have proved that, to the third order perturbation, the axion can be identified as the CDM in the general relativistic context. In our previous works we have shown that the leading nonlinear power spectra (of density and velocity perturbations) in Einstein's gravity (in the comoving gauge) give virtually identical results as in the Newtonian context \cite{JGNH-2011}. In the current paradigm of concordance cosmology the pure general relativistic contributions arising from the third order are entirely negligible in all scales compared with the relativistic/Newtonian results which are identified to the second order perturbation \cite{Noh-Hwang-2004}.

Using Eqs.\ (\ref{kappa-u}) and (\ref{X-Y-def}), Eq.\ (\ref{v-vector}) becomes
\bea
   & & \widehat v_i^{(v)}
       = - {\hbar^2 \over 4 c^2 m^2 a^2} \big\{
       \left( \nabla_i \delta \right) {\bf u} \cdot \nabla \delta
       - \nabla_i \Delta^{-1} \nabla \cdot
       \left[ \left( \nabla \delta \right) {\bf u}
       \cdot \nabla \delta \right] \big\}.
   \label{axion-rotation}
\eea
This is the third-order vector-type perturbation generated from the axion. From Eqs.\ (\ref{Tab-fluid}), (\ref{u_i}) and (\ref{v-decomposition}), to the third order, we have $\widetilde T^0_i / \mu = \widehat v_i^{(v)} = \widetilde u_i / a$. Thus, we may identify
\bea
   & & u_i = \nabla_i u + \widehat v_i^{(v)}.
\eea
In case of the zero-pressure medium like the CDM, the nonlinear order vector-type velocity (rotational) perturbation $\widehat v_i^{(v)}$ is {\it not} generated from the scalar-type perturbation, and the homogeneous equation has decaying nature ($\widehat v_i^{(v)} \propto 1/a$) in an expanding medium. This can be proved as the following. For a zero-pressure fluid in the comoving gauge, from the covariant momentum conservation in Eq.\ (3.9) of \cite{Noh-2014}, we have
\bea
   & & ( a {\widehat v}_i^{(v)} )^{\displaystyle\cdot}
       + c^2 \delta {\cal N}_{,i}
       = {\rm nonlinear \; terms \; all \; involving \;} \widehat v_i^{(v)},
\eea
valid to the fully nonlinear order. To the linear order, we have $\delta {\cal N} = 0$ and $\widehat v_i^{(v)} \propto 1/a$. As the $\widehat v_i^{(v)}$ is a homogeneous and pure decaying mode we may set $\widehat v_i^{(v)} \equiv 0$; in fact, as we consider a vector-type perturbation generated from the pure scalar-type perturbation, we may ignore the linear order $\widehat v_i^{(v)}$. This can be continued to all higher order perturbations with $\delta {\cal N} = 0$ and $\widehat v_i^{(v)} \propto 1/a$; as the $\widehat v_i^{(v)}$ is homogeneous (not sourced by the scalar-type perturbation) and pure decaying mode, we may set $\widehat v_i^{(v)} \equiv 0$ to all orders in perturbation. [End of the proof.] Notice that in the matter-dominated era with $K = 0 = \Lambda$, we have $\delta \propto a$ and ${\bf u} \propto a^{1/2}$, thus the axion-generated vector-type perturbation behaves as $\widehat v_i^{(v)} \propto a^{1/2}$.

%
%
\section{Discussion}
                                  \label{sec:discussion}

In this work we show that, in the super-Jeans scale, the axion behaves as the CDM to the third order perturbation. The axion pressure terms are important only in the cosmologically negligible scale for the canonical mass axion. In the axion-comoving gauge we have taken, in the super-Jeans scale, there exists the relativistic/Newtoninian correspondence to the second-order perturbations, and pure relativistic corrections appearing from the third-order are numerically negligible (in the current paradigm of concordance cosmology) \cite{JGNH-2011}. Therefore, in those scales (including the super-horizon scale) the axion is indistinguishable from the CDM as a zero-pressure fluid.

Concerning the axion pressure term, we show that our purely {\it classical} perturbation treatment of the axion as a massive scalar field in Einstein's gravity gives an {\it identical} result from quantum mechanical treatment in the non-relativistic limit of the Bose-Einstein condensation \cite{BEC,Sikivie-Yang-2009}. Although our result is valid only to third order perturbation, as the analysis is made in fully general relativistic context, the coincidence implies that the axion is non-relativistic indeed. In the regime where the axion pressure term becomes important, say for a extremely low mass axion where the axion Jeans scale becomes cosmologically relevant, besides the axion pressure term appearing from the linear order, the pure general relativistic contributions starting to appear from the third order through $\varphi$ also have axion pressure contribution with decaying nature though, see Eq.\ (\ref{varphi-axion}).

We also show that the axion supports the vector-type velocity perturbation from the third order whereas the pressureless fluid without the vector-type velocity perturbation to the linear order does not generate the vector-type velocity perturbation to the nonlinear order.

Our result includes the cosmological constant. Extending our analysis for a realistic cosmological situation in the presence of other components of fluids and fields is trivial. We emphasize that the spatial and temporal gauge condition in Eqs.\ (\ref{spatial-gauge}) and (\ref{axion-CG}) and the special identification of perturbed velocity variable in Eq.\ (\ref{kappa-u}) were essential to get the above equations.

One missing contribution we ignored in this work is the simultaneously excited tensor-type (transverse-tracefree) perturbation. In our previous works \cite{Third-2005,JGNH-2011} we have ignored the ${\bf Y}$-term which comes from the vector-type perturbation generated by the nonlinear scalar-type perturbation. In the same sense the tensor-type perturbation is also generated by the nonlinear scalar-type perturbation. These scalar-generated vector- and tensor-type perturbations start to have their roles from the third-order perturbation, and their quantitative effects will be studied in a later work \cite{HJN-2015}.

Whether the axion behaves as a CDM fluid to the fully nonlinear order in Einstein's gravity is left for future study; the fully nonlinear perturbation equations of the zero-pressure irrotational fluid in Einstein's gravity are presented in Eqs.\ (56)-(60) in \cite{fNL-2013}, and our analysis up to Eq.\ (\ref{T_0i}) is valid to fully nonlinear order.

%
%
\section*{Acknowledgments:}

H.N.\ was supported by National Research Foundation of Korea funded by the Korean Government (No.\ 2015R1A2A2A01002791).
J.H.\ was supported by Basic Science Research Program
through the National Research Foundation (NRF) of Korea funded by the Ministry of Science, ICT and future Planning (No. 2013R1A2A2A01068519).
C.G.P.\ was supported by Basic Science Research Program through the National Research Foundation of Korea (NRF) funded by the Ministry of Science, ICT and Future Planning (No.\ 2013R1A1A1011107).

%
%

\end{document}